\documentclass[usenatbib]{mnras}

\usepackage[T1]{fontenc}
\usepackage{ae,aecompl}

\usepackage{graphicx}      
\usepackage{epstopdf}
\usepackage{amsmath}    
\usepackage{amssymb}    
\usepackage{booktabs}
\usepackage{bm}

\newcommand{\beq}{\begin{equation}}
\newcommand{\eeq}{\end{equation}}

\newcommand{\Ts}{T_\mathrm{s}}
\newcommand{\Tb}{T_\mathrm{b}}
\newcommand{\rhob}{\rho_\mathrm{b}}

\newcommand{\Msun}{\mathrm M\odot}

\title[Cooling of neutron stars]{
Cooling of neutron stars with diffusive envelopes}

\author[M. V. Beznogov et al.]{
M. V. Beznogov$^{1}$\thanks{E-mail: \href{mailto:mikavb89@gmail.com} {mikavb89@gmail.com}},
M. Fortin$^{2}$,
P. Haensel$^{2}$,
D. G. Yakovlev$^{3}$,
J. L. Zdunik$^{2}$
\\
$^{1}$St~Petersburg Academic University, 8/3 Khlopina st., St~Petersburg 194021, Russia \\
$^{2}$Nicolaus Copernicus Astronomical Center,  Bartycka 18, Warsaw 00-716, Poland \\
$^{3}$Ioffe Institute, 26 Politekhnicheskaya st., St~Petersburg 194021, Russia \\
}

\date{Accepted . Received ; in original form}
\pubyear{2016}

\begin{document}
\label{firstpage}
\pagerange{\pageref{firstpage}--\pageref{lastpage}}
\maketitle

\begin{abstract}
We study the effects of heat blanketing envelopes of neutron stars
on their cooling. To this aim, we perform cooling simulations using
newly constructed models of the envelopes composed of binary ion
mixtures (H--He, He--C, C--Fe) varying the mass of lighter ions (H,
He or C) in the envelope. The results are compared with those
calculated using the standard models of the envelopes which contain
the layers of lighter (accreted) elements (H, He and C) on top of
the Fe layer, varying the mass of accreted elements. The main effect is that the
chemical composition of the envelopes influences their thermal 
conductivity and, hence, thermal insulation of the star. For
illustration, we apply these results to estimate
the internal temperature of the Vela pulsar and to study the
cooling of neutron stars of ages of $10^5-10^6$ yr at the photon
cooling stage. The uncertainties of the cooling models associated
with our poor knowledge of chemical composition of the heat
insulating envelopes strongly complicate theoretical reconstruction
of the internal structure of cooling neutron stars from observations of
their thermal surface emission.
\end{abstract}


\section{Introduction}
\label{sec:Intro}

The long-standing problem of modern studies of neutron stars is to
investigate the properties of superdense matter in their cores. One
of a few methods to achieve this goal is to study the thermal evolution
of neutron stars (particularly, cooling of isolated stars) and
compare the theoretical models with the available observational data (e.g.,
\citealt{YP04,Page_etal09}; \citealt*{PPP15} and references therein).
Different models of superdense matter predict different rates of neutrino
cooling of  neutron star interiors and, therefore, different surface temperatures as they are directly related to the internal temperatures. This allows one to select 
most suitable models of superdense matter from observations. It is a
challenging task for many reasons, but we mainly focus on one of
them. In order to explore the properties of superdense matter one
needs to calculate (constrain) the internal temperature of the star
from observations of its thermal surface emission.

We will restrict ourselves to not too young neutron stars (of age
$t\gtrsim 10-100$ yr) which are thermally equilibrated and
isothermal inside. The initial internal equilibration mainly
consists in the equilibration between the crust and the core of a star
due to distinctly different microphysics there (e.g., \citealt{Lattimer94,YKGH01}
and references therein). After the equilibration
the main temperature gradient in these stars
still persists in a thin outer heat blanketing envelope with rather
poor thermal conduction. It is usually sufficient to assume that
this envelope extends from the atmosphere bottom to the layer of the
density $\rhob \sim 10^{10}$~g~cm$^{-3}$. Its thickness does
not exceed a few hundred meters, and its mass is
$\lesssim 10^{-8}-10^{-7}$ M$\odot$. On the other hand, 
its mass cannot be smaller than the mass of the neutron star
atmosphere (typically $\sim 10^{-18}-10^{-16}$ $\Msun$).
Let $\Tb$ be the temperature at the bottom density $\rho=\rhob$.
Although strong magnetic fields in the envelopes of neutron stars
can affect the insulating properties of the envelopes and create
an anisotropic distribution of the effective surface temperature $\Ts$ 
(e.g. \citealt{PYCG03,PPP15}), we neglect such effects in the
present paper. Our results cannot be used to study the thermal structure
and evolution of neutron stars with very strong fields,
particularly, magnetars.
Even in this case the relation between $\Ts$ and $\Tb$,
required for cooling simulations and data analysis, is uncertain,
mainly because of uncertain thermal conduction in the blanketing
envelopes due to their unknown chemical composition.

Although the chemical composition of the heat blanketing envelopes is
really uncertain, there are some natural limitations (see, e.g.,
\citealt{PCY97} and references therein). For instance, at high temperatures $T$
and/or densities $\rho$ hydrogen transforms into helium due to thermo- or
pycno-nuclear reactions and beta captures. Approximately, this happens
at $T \gtrsim 4 \times 10^7$ K and/or $\rho \gtrsim 10^7$ g~cm$^{-3}$.
At higher $T \gtrsim 10^8$ K and/or
$\rho \gtrsim 10^9$ g~cm$^{-3}$, helium, in its turn, transforms into carbon. 
At still higher $T \gtrsim 10^9$ K and/or $\rho \gtrsim 10^{10}$ g~cm$^{-3}$ 
carbon transforms into heavier elements.

The standard model of heat blanketing
envelopes is the model developed by \citet*{PCY97} and elaborated
by \citet{PYCG03}. In this model (hereafter, the PCY97 model), the envelope is onion-like,
composed of  shells of pure elements with abrupt
boundaries between the shells. The number of  the shells is determined
by a single parameter, $\Delta M$,  the accumulated mass of light elements (H, He, C).
The width of each shell is limited
by nuclear transformations of light elements. In the absence of light elements,
the envelope is purely iron. In the presence of a thick layer of light elements,
the envelope consists of consecutive shells of hydrogen, helium,
carbon and iron. Although this model has proved to be useful,
it relies on the assumption of abrupt boundaries
between the layers of different
chemical species and employs a specific dependence of the composition on the
accumulated mass (governed by nuclear reactions and beta-captures).

Recently we have developed new models of diffusive heat blanketing
envelopes  \citep*{BPY16} which consist of binary
ion mixtures (either H--He, or He--C, or C--Fe) with a variable mass $\Delta M$
of  light elements (either H, or He or C, respectively).
They can be diffusively equilibrated or not. Since the ions have
a tendency for separation, such an envelope consists of
a top layer of lighter ions, a bottom layer of heavier ions, and a transition
layer in between. While considering the equilibrated envelopes
we have used proper relations for the ion diffusive currents taking
into account the Coulomb coupling of the ions and the presence
of temperature gradients.  Note that \citet{BPY16}
have also introduced a characteristic transition density $\rho^*$ 
from a lighter element to a heavier one instead of the mass of light
elements $\Delta M$. There is a one to one correspondence between
$\Delta M$ and $\rho^*$.

Let us stress that according to \citet{BPY16} the separation between the H and He ions,
as well as between the C and Fe ones is rather strong (due to the gravitational force)
leading to a narrow separation layer.
In contrast, the separation between  the He and C ions (which have almost the same
charge-to-mass ratios) is slower (due to the Coulomb force), resulting in a wider transition layer. In any case, even at not very small deviations from diffusive equilibrium,
the $\Ts-\Tb$ relation is rather insensitive to the structure
of the transition zone and depends almost solely on $\Delta M$. The
$\Ts-\Tb$ relations have been calculated assuming different
values of $\rhob=10^8$, $10^9$ and $10^{10}$ g~cm$^{-3}$ (which
are suitable for different cooling problems; see \citealt{BPY16}). For 
convenience of using in the computer codes, numerical results
have been approximated by analytic expressions. To visualize
the properties of the new heat blanketing envelopes, \cite{BPY16} 
plotted (their figs. 1--7) representative profiles of particle 
fractions and temperatures in the envelopes as the functions of the
density  as well as the appropriate $\Ts -\Tb$ and
$\Tb - \Delta M$ relations. Note that those
figures corresponded to a neutron star model with mass $M=1.4\,\Msun$ 
and radius $R=10$ km; the same model will be used to 
interpret the observations of the Vela pulsar in Section \ref{sec:Vela}.   

Here we apply new envelope models to simulate cooling of the isolated
middle-aged ($t \sim 10^2 - 10^6$ yr) neutron stars.
We will use the PCY97 model as a reference for comparison.
In Section \ref{sec:Formation} we present some formation scenarios
of accreted envelopes. In Sections \ref{sec:Vela}--\ref{sec:photon-cool} we outline some
calculations of internal structure and cooling of neutron stars with new envelope models.
Our conclusions are formulated in Section~\ref{sec:Concl}.

\section{Formation scenarios}
\label{sec:Formation}

The composition of heat blanketing envelopes of neutron stars is uncertain.
It can greatly vary depending on the formation history of the star
and its evolutionary scenario.

Initially, it was thought that the
envelopes (as well as neutron star atmospheres) contain heavy elements
like iron, as a result of the formation of the envelope in a hot and very young star where
light elements  burnt-out into heavier ones.
However, a more detailed analysis of  the observed spectra of the thermal radiation
originating from the atmospheres of neutron stars has shown that some spectra are
better described by blackbody models (of iron atmosphere models) while
others are better approximated by hydrogen atmosphere models
(see, e.g., \citealt{Potekhin_14} and references therein).
Moreover, the spectrum of the neutron star in the Cas A supernova remnant is
well described by a carbon atmosphere model \citep{HoHeinke_09}.
The same is true for the neutron star in
the supernova remnant HESS J1731--347 \citep{Klochkov_etal13}.

Therefore, the surface composition of thermally emitting
neutron stars can be different. Naturally, the composition of the underlying
envelopes can also be different. The composition of the surface layers
can be affected by the fallback of material on the neutron star surface
after the supernova explosion, by the accretion of hydrogen and helium from
interstellar medium or from a binary companion (if the neutron star
is or was in a compact binary), by the ion diffusion and the nuclear
evolution in the neutron star envelope, and by other effects.
A neutron star can directly accrete hydrogen and helium
(e.g., \citealt{BBMY92}). Alternatively, helium can be
produced in nuclear reactions after the accretion of hydrogen. Helium
can also burn further into carbon (see, e.g.,
\citealt{Rosen68,CB10}). In some cases the reverse process of
spallation of heavier elements into lighter ones is possible.
There are also indications that some transiently accreting neutron
stars in low mass X-ray binaries in quiescent states have their
outer envelopes composed of H and He which are left after an active
accretion phase \citep{BBC02}.

All in all, the composition of neutron star envelopes is largely
unknown. It seems instructive to consider different models of the envelopes
and to analyse observational manifestations of such models.

\section{Thermal state of the Vela pulsar}
\label{sec:Vela}

\begin{figure*}
    \centering
\includegraphics[height=8.0cm,keepaspectratio=true,clip=true,trim=0cm 0.2cm 0.3cm 0.8cm]{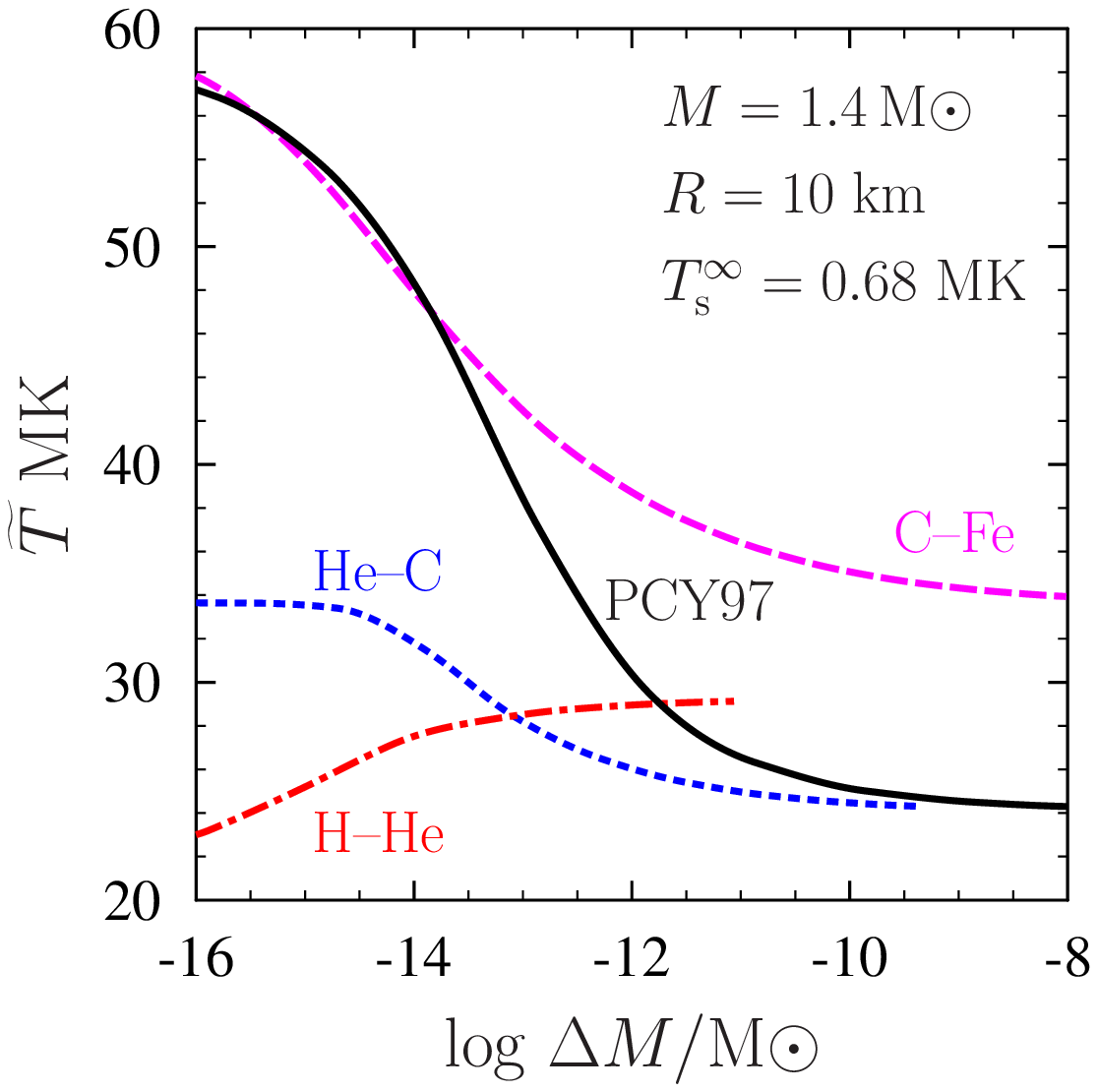}%
\includegraphics[height=8.0cm,keepaspectratio=true,clip=true,trim=0cm 0.2cm 0.3cm 0.8cm]{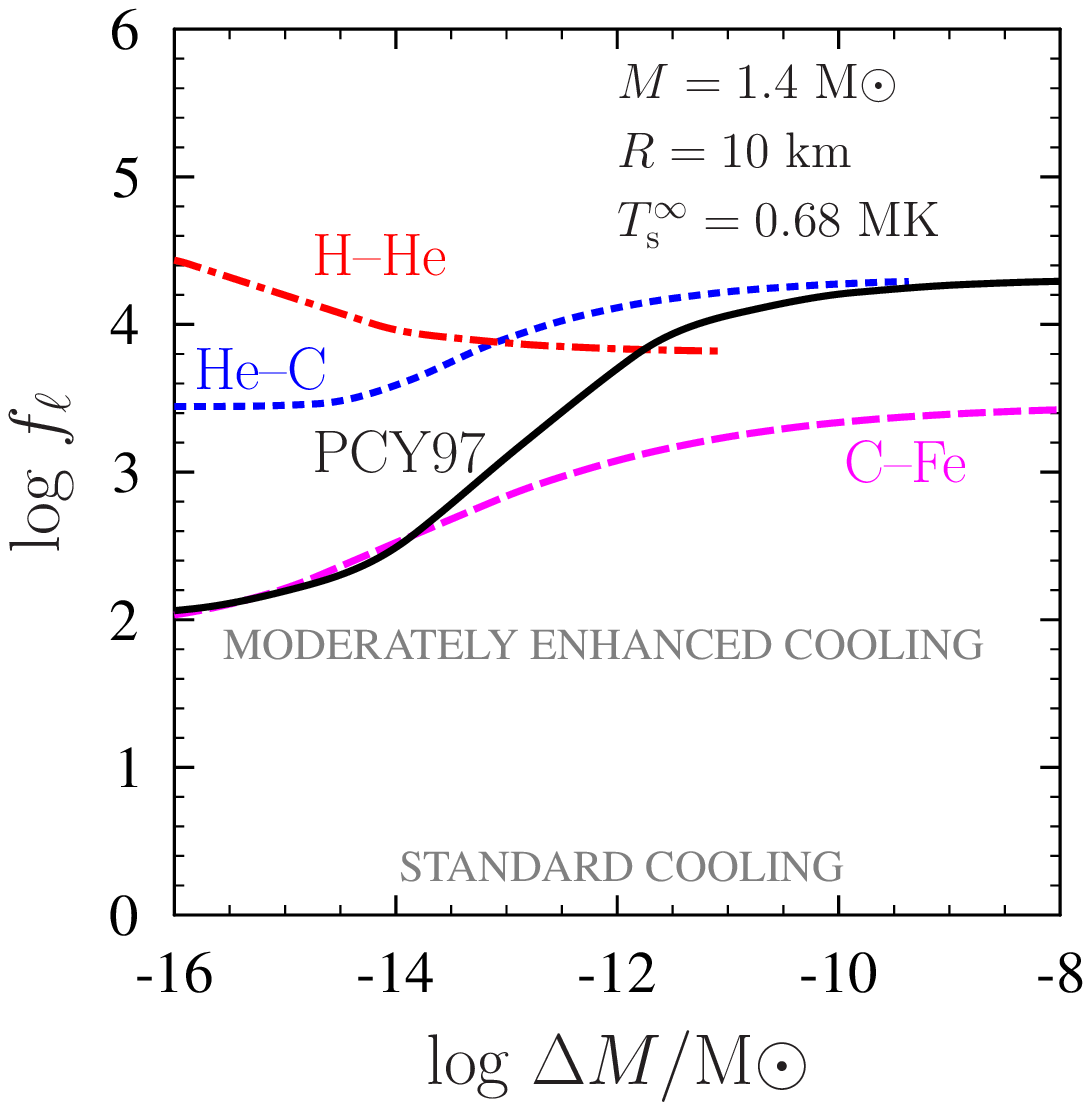}
\caption{Internal temperature $\widetilde{T}$ (left-hand panel) and
neutrino cooling function $f_\ell$ (right-hand panel) of the
Vela pulsar for H--He, He--C, and C--Fe models of heat blanketing
envelopes versus mass $\Delta M$ of the lighter element in the binary
mixture. For comparison, we present also  $\widetilde{T}$
and $f_\ell$ obtained using the envelope model of \citet{PCY97}
(PCY97) versus the mass $\Delta M$ of accreted elements. The level
of the standard neutrino cooling $f_\ell = 1$ on the right-hand panel
refers to a non-superfluid star which cools via the modified Urca
process. The level $f_\ell \approx 10^2$ of moderately enhanced neutrino
cooling can be provided by neutrino emission due to moderately
strong neutron superfluidity in the core. See text for details.}
\label{fig:tbvela}
\end{figure*}

\begin{figure*}
\centering
\includegraphics[height=8.0cm,keepaspectratio=true,clip=true,trim=0cm 0.2cm 0.3cm 0.8cm]{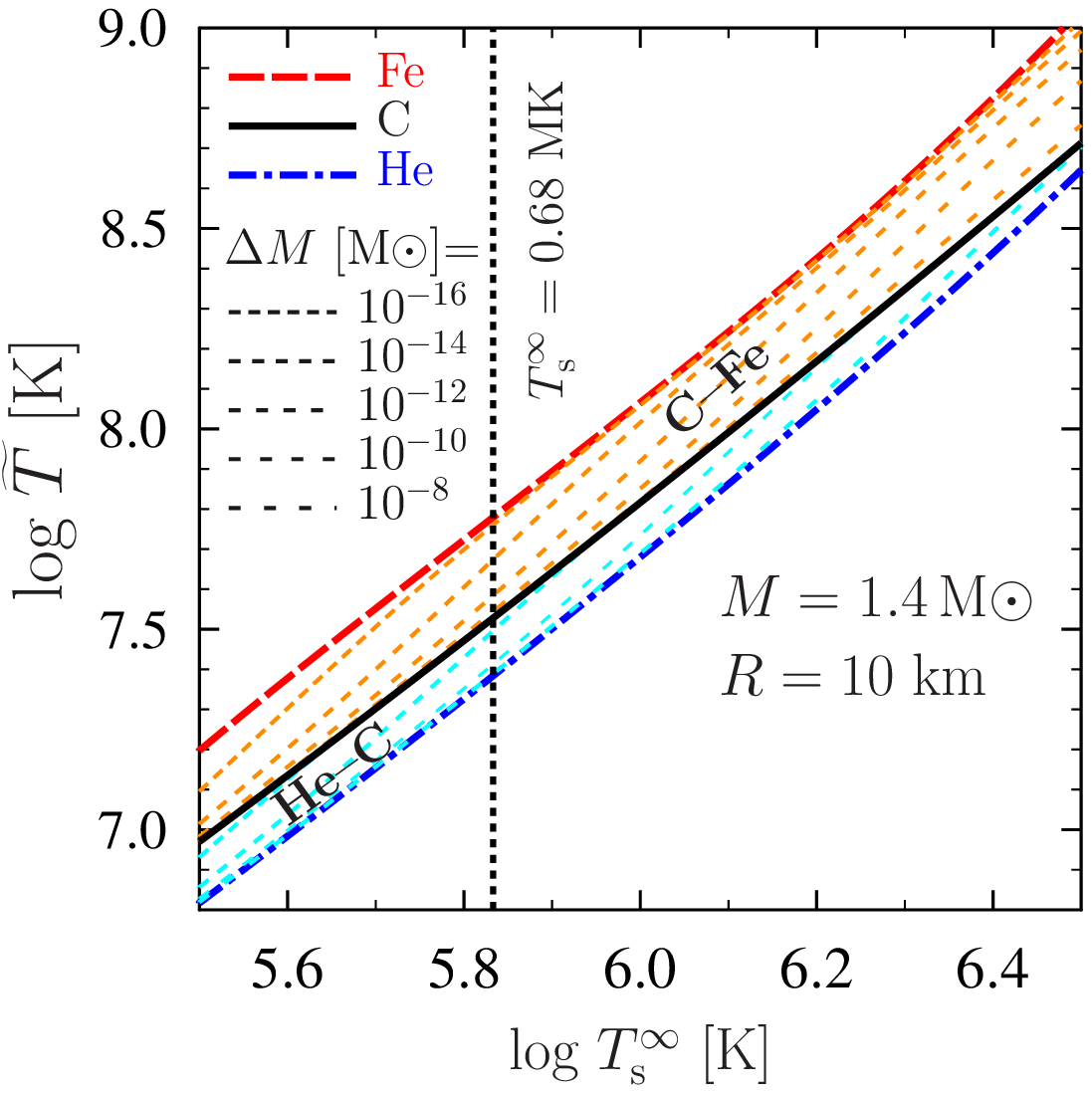}%
\includegraphics[height=8.0cm,keepaspectratio=true,clip=true,trim=0cm 0.2cm 0.3cm 0.8cm]{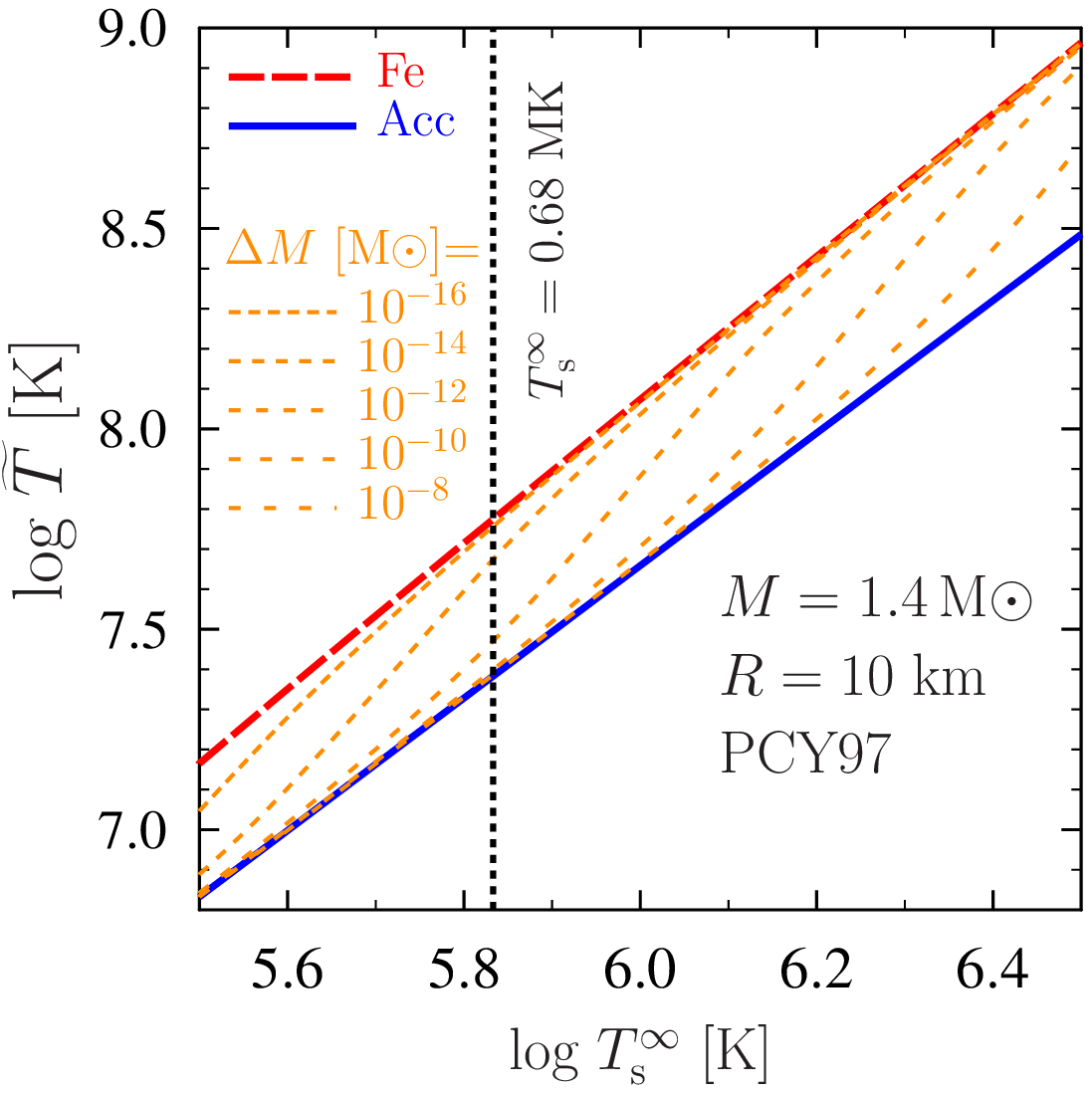}
\caption{Thermal states ($\widetilde{T}$ versus $\Ts^\infty$)
of the Vela pulsar with $M=1.4\,\Msun$ and $R=10$ km in the past
and the future (on the right and left of the vertical dotted lines,
respectively) for different chemical compositions of heat blanketing
envelopes. Left-hand panel: He--C and C--Fe envelopes. The thick lines
correspond from top to bottom to pure Fe, C and He envelopes,
respectively. The thinner different dashed lines are for binary mixtures
with different mass $\Delta M$ of lighter elements. Right-hand
panel: the PCY97 envelope.  The thick lines are for pure Fe and pure
accreted matter;  the thinner dashed lines are for different mass of the
accreted matter.
For given envelope models at a fixed $\Ts^\infty$ the internal
temperature $\widetilde{T}$ increases when $\Delta M$ decreases. }
\label{fig:tbtsvela}
\end{figure*}

Having a theoretical $\Ts-\Tb$ relation and the values
of $\Ts$ inferred from observations one can calculate the
internal temperature of a neutron star.

Let us illustrate this by taking the Vela pulsar as an example. It is
a middle-aged pulsar (with the characteristic pulsar age $\approx 11$ kyr).
Its internal thermal relaxation occurred long ago, so that its
redshifted internal temperature $\widetilde{T}$ is constant over the
pulsar interior (excluding the heat blanketing envelope). Using the
magnetic hydrogen atmosphere model and taking a gravitational mass $M=1.4\,\mathrm{M}\odot$ and a circumferential radius of
the star $R=10$ km  [with an apparent radius $R^\infty=R/\sqrt{1-x_\mathrm{g}}
=13$ km, where $x_\mathrm{g}=2GM/(Rc^2)$] \citet{Pavlov_etal01}
inferred the redshifted effective surface temperature 
$\Ts^\infty=0.68 \pm 0.03$ MK (at 68 per cent confidence level).
To be specific, we employ $\Ts^\infty= 0.68$ MK which corresponds to
$\Ts=\Ts^\infty/\sqrt{1-x_\mathrm{g}}=0.888$ MK. Using a
$\Ts-\Tb$ relation we can immediately calculate $\Tb$ 
and the redshifted internal temperature $\widetilde{T}=\Tb 
\sqrt{1-x_\mathrm{g}}$. Note that the Vela pulsar possesses a
magnetic field $B \sim 3 \times 10^{12}$ G, while we, following
\citet{BPY16}, neglect the effects of magnetic fields on the heat
blanketing envelope (as already mentioned above). We do it for simplicity and
illustration. In addition, such magnetic fields do not affect
strongly the $\Ts-\Tb$ relations \citep{PYCG03}.

The left-hand panel of Fig. \ref{fig:tbvela} shows the inferred
internal Vela's temperature $\widetilde{T}$ determined  for a number
of envelope models. The short-dashed line corresponds to $\widetilde{T}$ for
the He--C envelope with $\rhob=10^{10}$ g~cm$^{-3}$ as a
function of $\Delta M = \Delta M_\mathrm{He}$. The long-dashed line is
the same for the C--Fe envelope versus mass of carbon, $\Delta
M=\Delta M_\mathrm{C}$. The dot-dashed line is for the H--He envelope
with $\rhob=10^{8}$ g~cm$^{-3}$ (at higher $\rhob$ He
starts to burn in pycno-nuclear reactions) versus mass of hydrogen,
$\Delta M=\Delta M_\mathrm{H}$. The line is extended only to $\Delta
M_\mathrm{H}\sim 10^{-11}\,\mathrm{M}\odot$ to which hydrogen can survive
in dense matter (e.g. Section \ref{sec:Intro}). Finally, the solid PCY97
curve shows $\widetilde{T}$ for the H--He--C--Fe envelope of
\citet{PCY97} versus the mass $\Delta M$ of `accreted' elements
(H+He+C). The line for the H--He envelope is plotted up to
$\Delta M \sim 10^{-11}\,\Msun$, for the He--C envelope -- up to
$\Delta M \sim 10^{-9}\,\Msun$, two other lines are plotted up to
$\Delta M = 10^{-8}\,\Msun$, an approximate value of
$\Delta M$ to which corresponding envelopes can survive; see Section \ref{sec:Intro}.

For larger $\Delta M$ we have a more heat transparent envelope with
smaller internal temperature $\widetilde{T}$ for the same surface
temperature $\Ts^\infty$.
An exception from this rule is provided by the H--He envelope where
the situation is inverted as explained by \citet{BPY16}. One can see
that the variations of $\widetilde{T}$ with
$\Delta M$, indeed, prevent the accurate determination of
$\widetilde{T}$ if the envelope composition is unknown. The
strongest variations are seen to occur
for the PCY97 model, which takes into account a wider range of
elements. In the case of binary mixtures, the variations become smaller,
and they are especially small for the H--He and He--C mixtures.

Since the Vela pulsar is at the neutrino cooling stage with an
isothermal interior, its internal temperature $\widetilde{T}$
determines (e.g., \citealt{Yakovlev_etal11,Weisskopf_etal11}) the
fundamental parameter of superdense matter in its core, which
is the neutrino cooling function
\begin{equation}
     \ell(\widetilde{T})=L_\nu^\infty (\widetilde{T}) /C(\widetilde{T}),
\label{eq:ell}
\end{equation}
where $L_\nu^\infty$ is the redshifted neutrino luminosity of the
star, and $C$ is its heat capacity; both quantities are mainly
determined by the star's core. The convenient unit of
$\ell(\widetilde{T})=\ell(\widetilde{T})_\mathrm{SC} \propto
\widetilde{T}^7$ is provided by the so-called standard neutrino candle.
It corresponds to a non-superfluid star which cools via the
modified Urca processes of neutrino emission.
\citet{Yakovlev_etal11} as well as \citet{Ofengeim_etal15} obtained analytic
approximations for $\ell(\widetilde{T})_\mathrm{SC}$ calculated for a
number of neutron star models with different masses and nucleonic equations
of state (EOSs) in the core. These approximations are universal
(almost independent of the EOS) and more or less equivalent
\citep{Ofengeim_etal15}. They permit a model-independent analysis of the thermal
states of neutron stars. Such an analysis has been performed
previously for the Crab pulsar \citep{Weisskopf_etal11}; for the
neutron star in the Cas A supernova remnant neglecting and including
its possible rapid cooling in the present epoch
\citep{Yakovlev_etal11,SY15}; and for the neutron star in the HESS
J1731--347 supernova remnant \citep{Ofengeim_etal15}.

Let us perform similar analysis for the Vela pulsar. Taking possible
values of $\widetilde{T}$ from the right-hand panel of Fig.\
\ref{fig:tbvela} we can reconstruct $\ell(\widetilde{T})$. Using
the theoretical relations derived by \citet{Yakovlev_etal11} or
\citet{Ofengeim_etal15} we assume that the neutrino cooling function
of the Vela pulsar behaves as $\ell(\widetilde{T}) \propto
\widetilde{T}^7$. Then we can determine $\ell(\widetilde{T})$ for any value
of $\widetilde{T}$ and find
\begin{equation}
    f_\ell=\ell(\widetilde{T})/\ell(\widetilde{T})_\mathrm{SC},
\label{eq:fl}
\end{equation}
which is the Vela's neutrino cooling function expressed in terms of standard
candles. This analysis is valid for a wide range of physical
scenarios including (i) the standard candle cooling, $f_\ell=1$; (ii)
slower cooling through nucleon-nucleon bremsstrahlung of neutrino
pairs if the modified Urca process is suppressed by strong neutron
or proton superfluidity ($0.01 \lesssim f_\ell<1$); (iii) faster
cooling via neutrino emission due to moderately strong triplet-state
Cooper pairing of neutrons in the core ($1< f_\ell \lesssim 10^2$).

It is well known that the Vela pulsar cools somewhat faster than the
standard candle (e.g., \citealt{PLPS04,YP04}) so that $f_\ell>1$.
The values of $f_\ell$ derived from the values of $\widetilde{T}$
are plotted on the right-hand panel of Fig.\ \ref{fig:tbvela} as a
function of $\Delta M$. In order to infer $f_\ell$ we have used
theoretical formulae from \citet{Ofengeim_etal15}, but the formulae
from \citet{Yakovlev_etal11} would give similar results. As on the
left-hand panel, the lines of different types refer to the different
models of heat insulation in the Vela's envelope. The dependence of
$f_\ell$ on the chemical composition of the envelope is seen to be
very strong. By increasing the mass of accreted matter in the PCY97
model to the maximum possible value $\Delta M \sim 10^{-8}$ M$\odot$
we increase $f_\ell$ from about $10^2$ to $2 \times 10^4$.
Similarly, increasing the mass of carbon in the C--Fe envelope we
vary $f_\ell$ from about $10^2$ to $2.5 \times 10^3$. Finally, by
increasing the mass of He in the He--C envelope or the mass of H in
the H--He envelope we can vary $f_\ell$ within about one decade
around $f_\ell \sim 10^4$.

These results indicate once more that the chemical composition of
the envelope is of great importance for studying the internal structure
of neutron stars. In addition, the results for the Vela pulsar allow
us to draw an important conclusion. Specifically, let us assume
that we wish to describe the cooling of isolated neutron stars using
the minimal cooling theory \citep{PLPS04,GKYG04}. In this theory,
neutron stars have a nucleon core, the powerful direct Urca process of
neutrino emission \citep{LPPH91} is forbidden, and the cooling
enhancement over the standard neutrino candle is provided by the
neutrino emission due to a moderately strong triplet-state Cooper
pairing of neutrons. The minimal cooling theory states that in this
case the enhancement is limited to $f_\ell \lesssim 10^2$. Within
the minimum cooling paradigm, according to the right-hand panel of
Fig.\ \ref{fig:tbvela}, the Vela pulsar cannot possess H--He or
He--C envelopes. Its envelope should be mostly composed of iron. Its
core should mainly contain moderately strong neutron superfluidity
to ensure the maximum neutrino emission enhancement $f_\ell \sim
10^2$. The Vela pulsar is the isolated
neutron star coldest for its age. Its cooling regime should be similar
to that of the Cas A neutron star if the current rapid cooling of the Cas A star is real
(\citealt{HH10,PPLS11,Shternin_etal11,Elshamouty_etal13}; see, however,
\citealt{PPSK13} for the alternative view on the rapid cooling of the Cas A
star). The Cas A neutron star is just younger but could become as cold as Vela
in about 10 kyr.

On the other hand, we can adopt another cooling theory which would allow for
the existence of stronger neutrino emission (e.g., \citealt{YP04}) in the Vela pulsar,
for instance, due to direct Urca process or due to similar processes
enhanced, for instance, by pion condensation. Then the composition
of the envelope would become again rather uncertain which would
strongly complicate theoretical analysis of the internal structure
of the Vela pulsar.

In Fig.\ \ref{fig:tbtsvela} we show thermal states of the Vela pulsar in the
past and the future for the same assumptions of $M=1.4\,\mathrm{M}\odot$
and $R=10$ km as in \citet{Pavlov_etal01} and for different models
of the envelopes from Fig.\ \ref{fig:tbvela}. The thermal states are characterized by
the values of $\widetilde{T}$ versus $\Ts^\infty$. The vertical dotted lines
refer to the present epoch ($t=11$ kr, $\Ts^\infty=0.68$ MK). The left-hand panel
corresponds to the C--Fe and He--C envelopes. The thick lines are for the envelopes of
pure elements : Fe (long-dashed line), C (solid line), and He (dash-dotted line).
The thin dashed lines with different dash separations refer to binary mixtures with
different masses of lighter elements, $\Delta M/\mathrm{M}\odot=10^{-16}$, $10^{-14}$,
$10^{-12}$, $10^{-10}$ and $10^{-8}$.  The lowest $\Delta M$ corresponds to a very
thin outer layer of lighter element while the largest $\Delta M$ to the envelopes with
a very thin bottom layer of heavier element. One can observe the
evolution of $\widetilde{T}(\Ts^\infty)$ with increasing $\Delta M$
from the  $\widetilde{T}(\Ts^\infty)$ dependence
for pure heavier to pure lighter element. The older the
star (the smaller $\Ts^\infty$) the smaller mass of lighter element affects
$\widetilde{T}$.

The right-hand panel of  Fig.\ \ref{fig:tbtsvela} shows the thermal
states of the Vela pulsar for the PCY97 envelopes. The upper thick line is again
for pure Fe while the lower line is for purely accreted matter. The thin lines refer
to different masses of accreted matter. Note that the normalized neutrino cooling
function $f_\ell$ (in units of standard candles) does not evolve in time as long as
$\ell(\widetilde{T})\propto \widetilde{T}^7$ but remains the same as plotted on the
right-hand panel of Fig.\ \ref{fig:tbvela}.

\section{Examples of cooling calculations}
\label{sec:Calculs-Res}

Detailed calculations of neutron star cooling with new envelope models
are outside the scope of this paper. Let us outline some selected results.

\begin{figure*}
    \centering
    \includegraphics[width=0.25\textwidth]{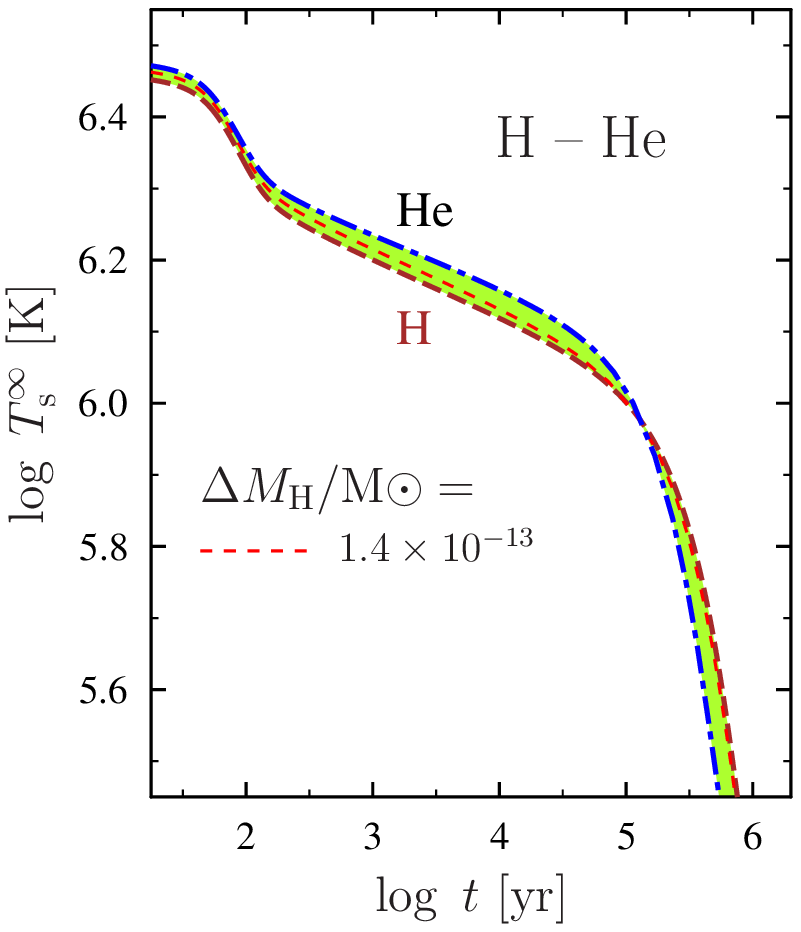}%
    \includegraphics[width=0.25\textwidth]{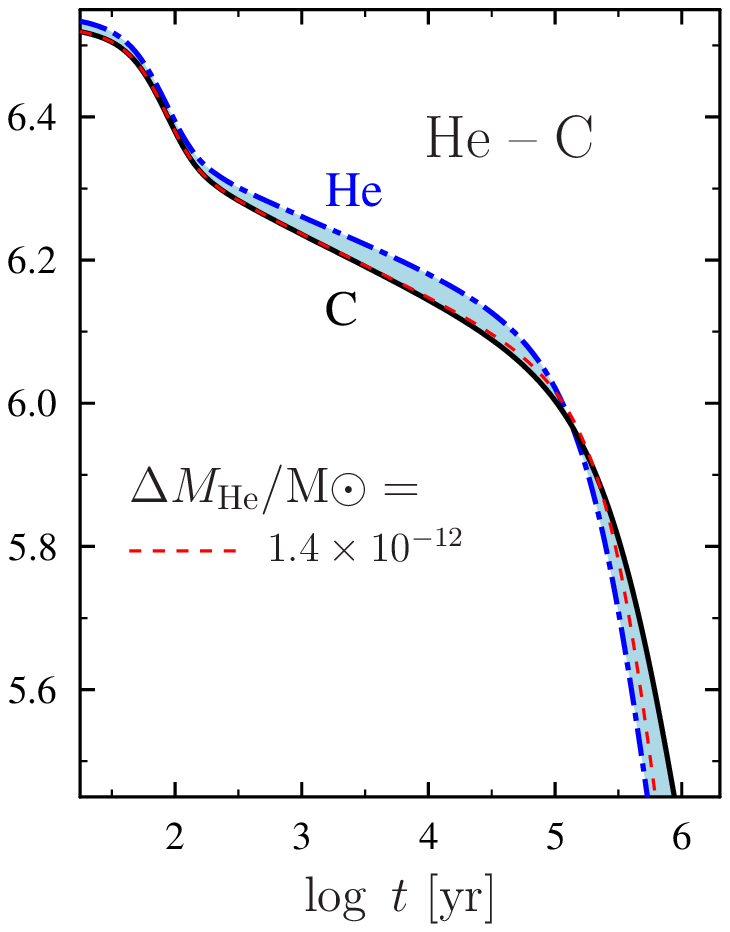}%
    \includegraphics[width=0.25\textwidth]{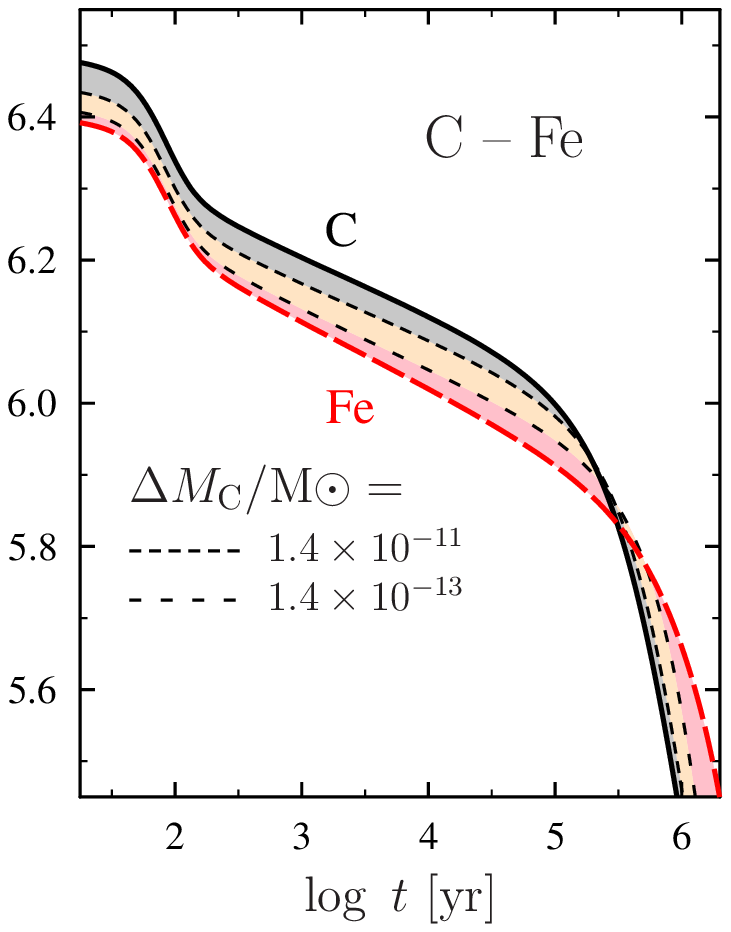}%
    \includegraphics[width=0.25\textwidth]{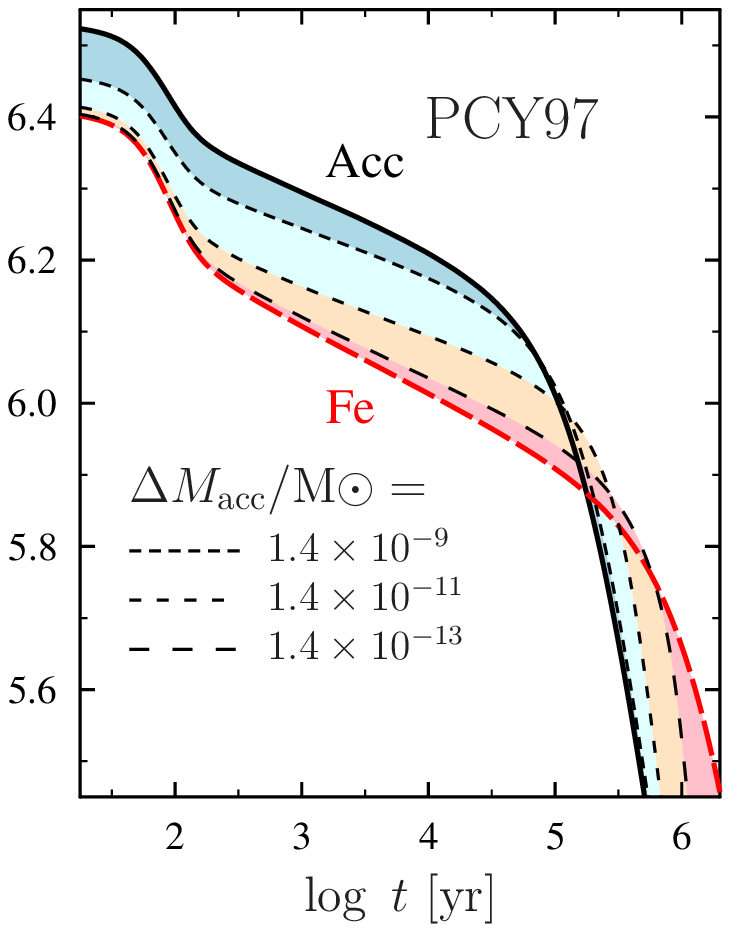}
    \caption{Cooling curves (redshifted effective surface temperature  $\Ts^\infty$
        versus age $t$) for a 1.4\,M$\odot$ non-superfluid neutron star with the BSk21 EOS, different chemical compositions of heat blanketing envelopes
        (from left to right: H--He, He--C, C--Fe, PCY97)
        and different accumulated masses of lighter elements. The
        curves for the envelopes with maximum $\Delta M$, containing 
        almost entirely lighter elements, correspond (from left to right) to $\Delta M_{\mathrm{H}} \sim 10^{-11}\,\Msun$, $\Delta M_{\mathrm{He}} \sim 10^{-9}\,\Msun$, $\Delta M_{\mathrm{C}} \sim 10^{-7}\,\Msun$
        and $\Delta M_{\mathrm{acc}} \sim 10^{-7}\,\Msun$, respectively.
        The curves for the envelopes with virtually no lighter elements are calculated
        for $\Delta M \sim 10^{-18}\,\Msun$. See text for details. }
    \label{fig:CoolCompar1.4}
\end{figure*}

To be specific, we restrict ourselves to the minimal cooling paradigm
\citep{PLPS04,GKYG04} mentioned above. Then the main regulators of neutron star cooling
are (i) the neutrino emission level, $f_\ell$ (which can vary from $\sim 10^{-2}$
to $\sim 10^2$ depending on superfluidity of neutrons and protons in the core) and (ii)
the composition of the envelope. Fixing $f_\ell$ and the envelope model but varying
$\Delta M$, we obtain a sequence of the cooling curves $\Ts^\infty(t)$. As a rule
these curves are almost `universal', independent of the EOS
in the core and of the star's mass $M$.
This is demonstrated, for instance, in figs.\ 24--26 of
\citet{YKGH01} for the case of standard neutrino candles ($f_\ell =1$, no superfluidity)
and iron envelopes.

For illustration, let us assume $f_\ell=1$ and focus on the effect of
the envelopes. Let us choose one neutron star model of mass $M=1.4\,$M$\odot$ with
the BSk21 EOS \citep{GCP10,PCGD12,Potekhin_etal13}. The stellar radius will then
be $R=12.60$ km, and the direct Urca process of neutrino emission will be forbidden.
The four panels of Fig.\ \ref{fig:CoolCompar1.4} show bands of the cooling curves
for such a star having different envelopes (from left to right: H--He, He--C, C--Fe,
and PCY97, respectively). Computations have been done using our general relativistic
cooling code \citep{GYP01} and cross checked with the `NScool'
cooling code by D. Page\footnote{`NScool' cooling code is available at
\url{http://www.astroscu.unam.mx/neutrones/NSCool/}.}.
The results of comparison are satisfactory:
calculated cooling curves differ slightly only at $t\gtrsim 2 $ Myr.
The initial segments of the cooling curves, $t \lesssim 10^2$ yrs refer
to the initial thermal relaxation within the star (e.g., \citealt{YP04}).

A band on each panel of  Fig.\ \ref{fig:CoolCompar1.4} is restricted
by upper and lower cooling curves (corresponding to almost pure He and H;
He and C; C and Fe; acc and Fe, respectively; `acc' refers to a
fully accreted PCY97 envelope). Short-dashed lines show some
intermediate cooling curves for a few values of $\Delta M$ to
demonstrate that the bands are actually filled by cooling curves
with different $\Delta M$.
Note that different binary mixtures are considered at different
$\rhob$ (the same as used in Section \ref{sec:Vela}).
Accordingly the He cooling curve for the H--He envelope is somewhat
different from the He curve for the He--C envelope.

Varying $\Delta M$ for the same envelope model, we change $\Ts^\infty(t)$. 
This can be treated as the `broadening' of the
cooling curve (because, as a rule, $\Delta M$ is unknown). As seen
from Fig.\ \ref{fig:CoolCompar1.4}, for the H--He and He--C
envelopes this broadening seems weak. However, for the C--Fe and
PCY97 envelopes (where the properties of various ion species,
particularly, their thermal insulation differ stronger) the
broadening is large and prevents the determination of
the internal temperature $\widetilde{T}$ from observations.

\section{Photon cooling stage}
\label{sec:photon-cool}

Another important feature of the cooling curves for different $\Delta M$ in
Fig.\ \ref{fig:CoolCompar1.4} is their inversion at certain $t$ when the
band of the curves becomes thin and then wider
again. The inversion is accompanied by the interchange of the cooling curves. For
instance, before the inversion on the right-hand panel the lowest cooling curve
is for the iron envelope while after the inversion the iron envelope produces the highest
cooling curve. The inversion epoch changes from $t\approx 10^5$ yr for the lighter
H--He and He--C envelopes to $(2-3) \times 10^5$ yr for the heavier C--Fe and PCY97
envelopes.

\begin{figure}
\centering
\includegraphics[height=8.0cm,keepaspectratio=true,clip=true,trim=0.05cm 0.15cm 1.2cm 1.1cm]{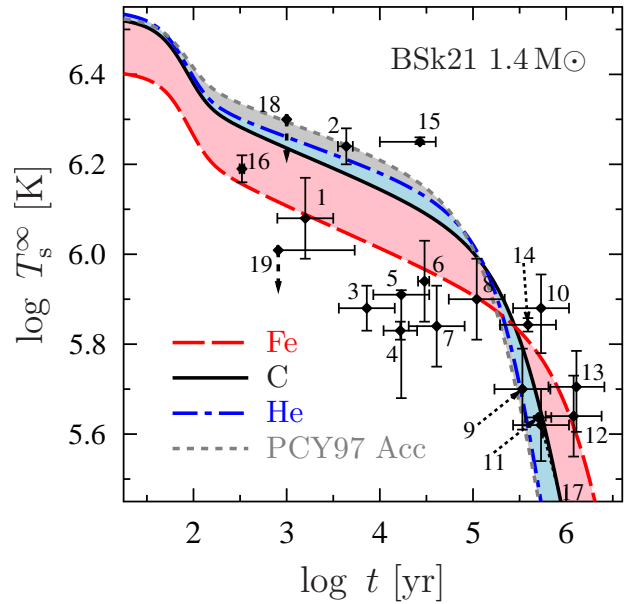}
\caption{Cooling curves for a 1.4\,M$\odot$ non-superfluid (standard
neutrino candle) neutron star with the BSk21 EOS. The thick lines
correspond  to envelopes of pure Fe, C, He, as well as of pure
accreted matter in the PCY97 model. The filled area between the Fe
and C curves is covered by cooling curves for C--Fe envelopes with
different $\Delta M$;  similarly,  the space between the C and He curves 
can be covered with cooling curves for He--C envelopes. The area
between the Fe and acc curves is covered by cooling curves for the
PCY97 envelopes. The cooling curves are compared with the observations
of isolated neutron stars. See the text for details.}
\label{fig:CoolCompar}
\end{figure}

These inversions are well known in the literature (e.g. \citealt{YP04}).
They manifest the transition from the neutrino cooling stage to the
photon cooling stage. The transition period is relatively short. The transition has
a dramatic impact on the cooling process. At the neutrino cooling stage, a star cools
via neutrinos from  the interior and looks colder for a more insulating envelope composed of
heavier elements. At the photon cooling stage, the star cools via photons from the surface.
The neutrino emission becomes insignificant for the cooling process, and the cooling
is governed by the heat capacity of the core and the heat transparency of the envelope.
More insulating envelopes of heavier elements produce hotter stars.

Fig.\ \ref{fig:CoolCompar} shows a selection of theoretical cooling
curves from Fig.\ \ref{fig:CoolCompar1.4} that are obtained for a
1.4\,M$\odot$ nonsuperfluid star with the BSk21 EOS in the core. The
four thick curves correspond to the envelopes made of pure Fe (the
long-dashed line), C (the solid line), He (the dot-dashed line) and
of pure accreted matter in the envelope model of PCY97 (the
short-dashed line). The space between the Fe and C curves is filled
by the cooling curves for the C--Fe envelope with different $\Delta
M$. The space between the C and He curves is filled by the cooling
curves for the He--C envelope. The space between the Fe and acc
curves is covered by the cooling curves for the PCY97 envelope. To
simplify Fig.\ \ref{fig:CoolCompar} we do not present the results
for the H--He envelope which can be easily visualized from the
left-hand panel of Fig. \ref{fig:CoolCompar1.4}.

In addition, Fig.\ \ref{fig:CoolCompar} presents the observational
data on the isolated middle-aged neutron stars. The data are the
same as those presented in \citet{BY15,BY15b,Ofengeim_etal15};
references to original publications can also be found there. Neutron
star labels are as follows,
(1) PSR J1119--6127;
(2) RX J0822--4300 (in Pup A);
(3) PSR J1357--6429;
(4) PSR B0833--45 (Vela);
(5) PSR B1706--44;
(6) PSR J0538+2817;
(7) PSR B2334+61;
(8) PSR B0656+14;
(9) PSR B0633+1748 (Geminga);
(10) PSR B1055--52 ;
(11) RX J1856.4--3754;
(12) PSR J2043+2740;
(13) RX J0720.4--3125;
(14) PSR J1741--2054;
(15) XMMU J1732--3445;
(16) Cas A neutron star;
(17) PSR J0357+3205 (Morla);
(18) PSR B0531+21 (Crab);
(19) PSR J0205+6449 (in 3C 58).

First let us outline briefly the sources which are at the neutrino cooling
stage ($t \lesssim 10^5$ yr). Recall that for a standard neutrino candle with
an iron envelope we would have one Fe (thick long-dashed ) cooling curve which
cannot explain many data. As seen from Fig.\ \ref{fig:CoolCompar}, even
varying the composition of the envelope of the standard candle we can explain much
more sources, although not all of them. The explanation of other sources at the
neutrino cooling stage would require deviations from the standard neutrino candle
(from $f_\ell=1$). For instance, the hottest XMMU J1732--3445 source
can be explained assuming nearly maximum amount of carbon in the envelope and
strong proton superfluidity in the core ($f_\ell \sim 0.01$,
\citealt{Klochkov_etal15,Ofengeim_etal15}). The coldest sources at the
neutrino cooling stage, like the Vela pulsar, can be interpreted, for instance,
as neutron stars with moderately strong triplet-state neutron pairing in the
core which increases the neutrino cooling level to $f_\ell \sim 10^2$
(Section \ref{sec:Vela}).

Now let us focus on the sources at the photon cooling stage (after
the inversion of the cooling curves). It is not a surprise (Fig.\
\ref{fig:CoolCompar}) that all these sources are compatible with the
standard neutrino cooling candle (see below). The variety of such
sources can be explained by different composition of their
envelopes. The hottest neutron stars at the photon cooling stage
(like PSR B1055--52, \citealt{PZ03}; PSR J2043+2740,
\citealt{Zavlin09}; RX J0720.4--3125, \citealt{MZH03}) should have
their envelope made predominantly of iron while the coldest stars
(like Geminga, \citealt{KPZR05};
 RX J1856.4--3754, \citealt{Ho_etal07,Potekhin_14}) may have envelopes
 of lighter elements (for instance, carbon).

Note that although the evolution of neutron stars at the photon
cooling stage does not depend directly on their neutrino emission,
the observed sources with $t \gtrsim 10^5$ yr should not have
$f_\ell \gg 1$.
Otherwise they would cool rapidly at the neutrino cooling stage,
transit to the photon cooling stage earlier and would become very
weak at $t \gtrsim 10^5$ yrs. Therefore, the hottest neutron stars
at the photon cooling stage seem to be those which have $f_\ell
\lesssim 1$ and possess Fe envelopes (which are better thermal
insulators than the envelopes of lighter elements).

The effects of heat blanketing envelopes on the cooling of the neutron stars
with different masses, radii, equations of state of superdense matter, magnetic
fields, and superfluid properties have been extensively 
studied using the PCY97 envelopes and their magnetic extensions
(see, e.g., \citealt{YP04,KGYG2006,PPP15,BY15} and
references therein). For instance, it is well known that all cooling
curves $\Ts^\infty(t)$ for non-superfluid neutron stars of 
different masses which cool via modified Urca process ($f_\ell=1$) and
have iron heat blanketing envelopes but no strong magnetic fields,
merge in almost one and the same cooling curve. The same is also true
for the stars with fully accreted PCY97 heat blanketing envelopes
although the curve becomes different.
We have checked that this property survives for the new envelope
models of \citet{BPY16}. 

The above analysis has neglected possible mechanisms of neutron star reheating, for instance,
due to ohmic decay of magnetic fields, possible violations of beta-equilibrium, etc.;
e.g., \citet{YP04,PGW06} and references therein. Were these
mechanisms operative they would be able to keep neutron stars warmer. No such mechanisms
seem to be required for ordinary cooling middle-aged neutron stars.

\section{Conclusions}
\label{sec:Concl}

We have outlined the effects of our new models
for heat blanketing envelopes \citep{BPY16}
of neutron stars on the cooling
and thermal structure of isolated middle-aged neutron stars.
The new envelopes are composed of binary ionic mixtures
(either H--He, or He--C, or C--Fe) with any allowed mass $\Delta M$
of lighter elements. The results are compared with the
standard PCY97 models of the envelopes containing shells
 of H, He, C, and Fe with any possible mass of `accreted'
(H+He+C) elements \citep{PCY97}. As discussed in \citet{BPY16},
the new models allow one to consider wider classes of the envelopes.

In Section \ref{sec:Formation} we have outlined some formation scenarios of the envelopes.
In Section \ref{sec:Vela} we have considered the effects of the envelopes
on inferring the internal temperatures $\widetilde{T}$
of neutron stars from
observations and on constraining their neutrino cooling
function $f_\ell$ (the fundamental parameter of superdense matter
in a neutron star core). We have taken
the Vela pulsar as an example. The results confirm previous conclusions
(e.g., \citealt{Yakovlev_etal11,Weisskopf_etal11,Klochkov_etal15,Ofengeim_etal15})
that the composition of the heat blanketing envelope
is a major ingredient for the correct interpretation of observations. The
uncertainty in the envelope composition translates into a factor
of $\sim 10^2$ uncertainty in $f_\ell$. Nevertheless, since the Vela pulsar is
sufficiently cold \citep{Pavlov_etal01}, we have been able to conclude that
within the minimal cooling scenario the pulsar should have $f_\ell \sim 10^2$ and
the envelope predominantly made of iron.

In Sections \ref{sec:Calculs-Res} and \ref{sec:photon-cool} we have performed
some cooling calculations for a 1.4\,M$\odot$ neutron star with different envelopes
and compared the results with observations. A special emphasis has been made
on neutron stars of ages $t\sim 0.1-1$ Myr which have changed their cooling
regime from the earlier neutrino cooling to the photon cooling. We have shown that
all observations of such stars (including PSR B1055--52,
PSR J2043+2740, RX J0720.4--3125, Geminga, and
RX J1856.4--3754) are consistent with the scenario
in which these stars were initially nearly standard
neutrino candles or slower neutrino coolers ($f_\ell \lesssim 1$)
but possess various envelopes mostly containing carbon and iron.

Our consideration is definitely not complete. More work is required to overcome
the problem of heat blanketing envelopes in the theory of thermal evolution of neutron stars.
In particular, more complicated models of the envelopes can be constructed
taking into account multicomponent ion mixtures in and out of diffusive equilibrium;
the dynamical evolution of the diffusive equilibrium can also be modelled. In addition, one can
elaborate the existing models of diffusive nuclear burning in the envelopes
(e.g., \citealt{CB03,CB04,CB10}) which is neglected here. It would also
be very important to include the effects of magnetic fields \citep{PPP15} on the
envelopes of various types. Any additional reliable information
on the formation history and evolution of the envelopes
would also be most welcome. However, all these problems go far beyond the scope
of the present investigation.

\section*{acknowledgements}
The work of MB was partly supported by the Dynasty Foundation, the
work of DG by the Russian Foundation for Basic Research (grants
14-02-00868-a and 16-29-13009-ofi-m), and the work of MF,
PH, and LZ was supported by the Polish NCN research
grant no. 2013/11/B/ST9/04528.

\bsp
\label{lastpage}

\begin{thebibliography}{}
	\makeatletter
	\relax
	\def\mn@urlcharsother{\let\do\@makeother \do\$\do\&\do\#\do\^\do\_\do\%\do\~}
	\def\mn@doi{\begingroup\mn@urlcharsother \@ifnextchar [ {\mn@doi@}
		{\mn@doi@[]}}
	\def\mn@doi@[#1]#2{\def\@tempa{#1}\ifx\@tempa\@empty \href
		{http://dx.doi.org/#2} {doi:#2}\else \href {http://dx.doi.org/#2} {#1}\fi
		\endgroup}
	\def\mn@eprint#1#2{\mn@eprint@#1:#2::\@nil}
	\def\mn@eprint@arXiv#1{\href {http://arxiv.org/abs/#1} {{\tt arXiv:#1}}}
	\def\mn@eprint@dblp#1{\href {http://dblp.uni-trier.de/rec/bibtex/#1.xml}
		{dblp:#1}}
	\def\mn@eprint@#1:#2:#3:#4\@nil{\def\@tempa {#1}\def\@tempb {#2}\def\@tempc
		{#3}\ifx \@tempc \@empty \let \@tempc \@tempb \let \@tempb \@tempa \fi \ifx
		\@tempb \@empty \def\@tempb {arXiv}\fi \@ifundefined
		{mn@eprint@\@tempb}{\@tempb:\@tempc}{\expandafter \expandafter \csname
			mn@eprint@\@tempb\endcsname \expandafter{\@tempc}}}
	
	\bibitem[\protect\citeauthoryear{{Beznogov} \& {Yakovlev}}{{Beznogov} \&
		{Yakovlev}}{2015a}]{BY15}
	{Beznogov} M.~V.,  {Yakovlev} D.~G.,  2015a, \mn@doi [MNRAS]
	{10.1093/mnras/stu2506}, \href
	{http://adsabs.harvard.edu/abs/2015MNRAS.447.1598B} {447, 1598}
	
	\bibitem[\protect\citeauthoryear{{Beznogov} \& {Yakovlev}}{{Beznogov} \&
		{Yakovlev}}{2015b}]{BY15b}
	{Beznogov} M.~V.,  {Yakovlev} D.~G.,  2015b, \mn@doi [MNRAS]
	{10.1093/mnras/stv1293}, \href
	{http://adsabs.harvard.edu/abs/2015MNRAS.452..540B} {452, 540}
	
	\bibitem[\protect\citeauthoryear{{Beznogov}, {Potekhin}  \&
		{Yakovlev}}{{Beznogov} et~al.}{2016}]{BPY16}
	{Beznogov} M.~V.,  {Potekhin} A.~Y.,   {Yakovlev} D.~G.,  2016, \mn@doi [MNRAS]
	{10.1093/mnras/stw751}, \href
	{http://adsabs.harvard.edu/abs/2016MNRAS.459.1569B} {459, 1569}
	
	\bibitem[\protect\citeauthoryear{{Blaes}, {Blandford}, {Madau}  \&
		{Yan}}{{Blaes} et~al.}{1992}]{BBMY92}
	{Blaes} O.~M.,  {Blandford} R.~D.,  {Madau} P.,   {Yan} L.,  1992, \mn@doi
	[ApJ] {10.1086/171955}, \href
	{http://adsabs.harvard.edu/abs/1992ApJ...399..634B} {399, 634}
	
	\bibitem[\protect\citeauthoryear{{Brown}, {Bildsten}  \& {Chang}}{{Brown}
		et~al.}{2002}]{BBC02}
	{Brown} E.~F.,  {Bildsten} L.,   {Chang} P.,  2002, \mn@doi [ApJ]
	{10.1086/341066}, \href {http://adsabs.harvard.edu/abs/2002ApJ...574..920B}
	{574, 920}
	
	\bibitem[\protect\citeauthoryear{{Chang} \& {Bildsten}}{{Chang} \&
		{Bildsten}}{2003}]{CB03}
	{Chang} P.,  {Bildsten} L.,  2003, \mn@doi [ApJ] {10.1086/345551}, \href
	{http://adsabs.harvard.edu/abs/2003ApJ...585..464C} {585, 464}
	
	\bibitem[\protect\citeauthoryear{{Chang} \& {Bildsten}}{{Chang} \&
		{Bildsten}}{2004}]{CB04}
	{Chang} P.,  {Bildsten} L.,  2004, \mn@doi [ApJ] {10.1086/382271}, \href
	{http://adsabs.harvard.edu/abs/2004ApJ...605..830C} {605, 830}
	
	\bibitem[\protect\citeauthoryear{{Chang}, {Bildsten}  \& {Arras}}{{Chang}
		et~al.}{2010}]{CB10}
	{Chang} P.,  {Bildsten} L.,   {Arras} P.,  2010, \mn@doi [ApJ]
	{10.1088/0004-637X/723/1/719}, \href
	{http://adsabs.harvard.edu/abs/2010ApJ...723..719C} {723, 719}
	
	\bibitem[\protect\citeauthoryear{{Elshamouty}, {Heinke}, {Sivakoff}, {Ho},
		{Shternin}, {Yakovlev}, {Patnaude}  \& {David}}{{Elshamouty}
		et~al.}{2013}]{Elshamouty_etal13}
	{Elshamouty} K.~G.,  {Heinke} C.~O.,  {Sivakoff} G.~R.,  {Ho} W.~C.~G.,
	{Shternin} P.~S.,  {Yakovlev} D.~G.,  {Patnaude} D.~J.,   {David} L.,  2013,
	\mn@doi [ApJ] {10.1088/0004-637X/777/1/22}, \href
	{http://adsabs.harvard.edu/abs/2013ApJ...777...22E} {777, 22}
	
	\bibitem[\protect\citeauthoryear{{Gnedin}, {Yakovlev}  \& {Potekhin}}{{Gnedin}
		et~al.}{2001}]{GYP01}
	{Gnedin} O.~Y.,  {Yakovlev} D.~G.,   {Potekhin} A.~Y.,  2001, \mn@doi [MNRAS]
	{10.1046/j.1365-8711.2001.04359.x}, \href
	{http://adsabs.harvard.edu/abs/2001MNRAS.324..725G} {324, 725}
	
	\bibitem[\protect\citeauthoryear{{Goriely}, {Chamel}  \& {Pearson}}{{Goriely}
		et~al.}{2010}]{GCP10}
	{Goriely} S.,  {Chamel} N.,   {Pearson} J.~M.,  2010, \mn@doi [Phys. Rev. C]
	{10.1103/PhysRevC.82.035804}, \href
	{http://adsabs.harvard.edu/abs/2010PhRvC..82c5804G} {82, 035804}
	
	\bibitem[\protect\citeauthoryear{{Gusakov}, {Kaminker}, {Yakovlev}  \&
		{Gnedin}}{{Gusakov} et~al.}{2004}]{GKYG04}
	{Gusakov} M.~E.,  {Kaminker} A.~D.,  {Yakovlev} D.~G.,   {Gnedin} O.~Y.,  2004,
	\mn@doi [A\&A] {10.1051/0004-6361:20041006}, \href
	{http://adsabs.harvard.edu/abs/2004A%26A...423.1063G} {423, 1063}
		
		\bibitem[\protect\citeauthoryear{{Heinke} \& {Ho}}{{Heinke} \&
			{Ho}}{2010}]{HH10}
		{Heinke} C.~O.,  {Ho} W.~C.~G.,  2010, \mn@doi [ApJL]
		{10.1088/2041-8205/719/2/L167}, \href
		{http://adsabs.harvard.edu/abs/2010ApJ...719L.167H} {719, L167}
		
		\bibitem[\protect\citeauthoryear{{Ho} \& {Heinke}}{{Ho} \&
			{Heinke}}{2009}]{HoHeinke_09}
		{Ho} W.~C.~G.,  {Heinke} C.~O.,  2009, \mn@doi [Nature] {10.1038/nature08525},
		\href {http://adsabs.harvard.edu/abs/2009Natur.462...71H} {462, 71}
		
		\bibitem[\protect\citeauthoryear{{Ho}, {Kaplan}, {Chang}, {van Adelsberg}  \&
			{Potekhin}}{{Ho} et~al.}{2007}]{Ho_etal07}
		{Ho} W.~C.~G.,  {Kaplan} D.~L.,  {Chang} P.,  {van Adelsberg} M.,   {Potekhin}
		A.~Y.,  2007, \mn@doi [MNRAS] {10.1111/j.1365-2966.2006.11376.x}, \href
		{http://adsabs.harvard.edu/abs/2007MNRAS.375..821H} {375, 821}
		
		\bibitem[\protect\citeauthoryear{{Kaminker}, {Gusakov}, {Yakovlev}  \&
			{Gnedin}}{{Kaminker} et~al.}{2006}]{KGYG2006}
		{Kaminker} A.~D.,  {Gusakov} M.~E.,  {Yakovlev} D.~G.,   {Gnedin} O.~Y.,  2006,
		\mn@doi [MNRAS] {10.1111/j.1365-2966.2005.09812.x}, \href
		{http://adsabs.harvard.edu/abs/2006MNRAS.365.1300K} {365, 1300}
		
		\bibitem[\protect\citeauthoryear{{Kargaltsev}, {Pavlov}, {Zavlin}  \&
			{Romani}}{{Kargaltsev} et~al.}{2005}]{KPZR05}
		{Kargaltsev} O.~Y.,  {Pavlov} G.~G.,  {Zavlin} V.~E.,   {Romani} R.~W.,  2005,
		\mn@doi [ApJ] {10.1086/429368}, \href
		{http://adsabs.harvard.edu/abs/2005ApJ...625..307K} {625, 307}
		
		\bibitem[\protect\citeauthoryear{{Klochkov}, {P{\"u}hlhofer}, {Suleimanov},
			{Simon}, {Werner}  \& {Santangelo}}{{Klochkov}
			et~al.}{2013}]{Klochkov_etal13}
		{Klochkov} D.,  {P{\"u}hlhofer} G.,  {Suleimanov} V.,  {Simon} S.,  {Werner}
		K.,   {Santangelo} A.,  2013, \mn@doi [A\&A] {10.1051/0004-6361/201321740},
		\href {http://adsabs.harvard.edu/abs/2013A%26A...556A..41K} {556, A41}
			
			\bibitem[\protect\citeauthoryear{{Klochkov}, {Suleimanov}, {P{\"u}hlhofer},
				{Yakovlev}, {Santangelo}  \& {Werner}}{{Klochkov}
				et~al.}{2015}]{Klochkov_etal15}
			{Klochkov} D.,  {Suleimanov} V.,  {P{\"u}hlhofer} G.,  {Yakovlev} D.~G.,
			{Santangelo} A.,   {Werner} K.,  2015, \mn@doi [A\&A]
			{10.1051/0004-6361/201424683}, \href
			{http://adsabs.harvard.edu/abs/2015A%26A...573A..53K} {573, A53}
				
				\bibitem[\protect\citeauthoryear{{Lattimer}, {Pethick}, {Prakash}  \&
					{Haensel}}{{Lattimer} et~al.}{1991}]{LPPH91}
				{Lattimer} J.~M.,  {Pethick} C.~J.,  {Prakash} M.,   {Haensel} P.,  1991,
				\mn@doi [Phys. Rev. Lett.] {10.1103/PhysRevLett.66.2701}, \href
				{http://adsabs.harvard.edu/abs/1991PhRvL..66.2701L} {66, 2701}
				
				\bibitem[\protect\citeauthoryear{{Lattimer}, {van Riper}, {Prakash}  \&
					{Prakash}}{{Lattimer} et~al.}{1994}]{Lattimer94}
				{Lattimer} J.~M.,  {van Riper} K.~A.,  {Prakash} M.,   {Prakash} M.,  1994,
				\mn@doi [ApJ] {10.1086/174025}, \href
				{http://adsabs.harvard.edu/abs/1994ApJ...425..802L} {425, 802}
								
				\bibitem[\protect\citeauthoryear{{Motch}, {Zavlin}  \& {Haberl}}{{Motch}
					et~al.}{2003}]{MZH03}
				{Motch} C.,  {Zavlin} V.~E.,   {Haberl} F.,  2003, \mn@doi [A\&A]
				{10.1051/0004-6361:20030802}, \href
				{http://adsabs.harvard.edu/abs/2003A%26A...408..323M} {408, 323}
					
					\bibitem[\protect\citeauthoryear{{Ofengeim}, {Kaminker}, {Klochkov},
						{Suleimanov}  \& {Yakovlev}}{{Ofengeim} et~al.}{2015}]{Ofengeim_etal15}
					{Ofengeim} D.~D.,  {Kaminker} A.~D.,  {Klochkov} D.,  {Suleimanov} V.,
					{Yakovlev} D.~G.,  2015, \mn@doi [MNRAS] {10.1093/mnras/stv2204}, \href
					{http://adsabs.harvard.edu/abs/2015MNRAS.454.2668O} {454, 2668}
					
					\bibitem[\protect\citeauthoryear{{Page}, {Lattimer}, {Prakash}  \&
						{Steiner}}{{Page} et~al.}{2004}]{PLPS04}
					{Page} D.,  {Lattimer} J.~M.,  {Prakash} M.,   {Steiner} A.~W.,  2004, \mn@doi
					[ApJS] {10.1086/424844}, \href
					{http://adsabs.harvard.edu/abs/2004ApJS..155..623P} {155, 623}
					
					\bibitem[\protect\citeauthoryear{{Page}, {Geppert}  \& {Weber}}{{Page}
						et~al.}{2006}]{PGW06}
					{Page} D.,  {Geppert} U.,   {Weber} F.,  2006, \mn@doi [Nucl. Phys. A]
					{10.1016/j.nuclphysa.2005.09.019}, \href
					{http://adsabs.harvard.edu/abs/2006NuPhA.777..497P} {777, 497}
					
					\bibitem[\protect\citeauthoryear{{Page}, {Lattimer}, {Prakash}  \&
						{Steiner}}{{Page} et~al.}{2009}]{Page_etal09}
					{Page} D.,  {Lattimer} J.~M.,  {Prakash} M.,   {Steiner} A.~W.,  2009, \mn@doi
					[ApJ] {10.1088/0004-637X/707/2/1131}, \href
					{http://adsabs.harvard.edu/abs/2009ApJ...707.1131P} {707, 1131}
					
					\bibitem[\protect\citeauthoryear{{Page}, {Prakash}, {Lattimer}  \&
						{Steiner}}{{Page} et~al.}{2011}]{PPLS11}
					{Page} D.,  {Prakash} M.,  {Lattimer} J.~M.,   {Steiner} A.~W.,  2011, \mn@doi
					[Phys. Rev. Lett.] {10.1103/PhysRevLett.106.081101}, \href
					{http://adsabs.harvard.edu/abs/2011PhRvL.106h1101P} {106, 081101}
					
					\bibitem[\protect\citeauthoryear{{Pavlov} \& {Zavlin}}{{Pavlov} \&
						{Zavlin}}{2003}]{PZ03}
					{Pavlov} G.~G.,  {Zavlin} V.~E.,  2003, in {Bandiera} R.,  {Maiolino} R.,
					{Mannucci} F.,  eds, Texas in Tuscany. XXI Symposium on Relativistic
					Astrophysics. pp 319--328
					
					\bibitem[\protect\citeauthoryear{{Pavlov}, {Zavlin}, {Sanwal}, {Burwitz}  \&
						{Garmire}}{{Pavlov} et~al.}{2001}]{Pavlov_etal01}
					{Pavlov} G.~G.,  {Zavlin} V.~E.,  {Sanwal} D.,  {Burwitz} V.,   {Garmire}
					G.~P.,  2001, \mn@doi [ApJL] {10.1086/320342}, \href
					{http://adsabs.harvard.edu/abs/2001ApJ...552L.129P} {552, L129}
					
					\bibitem[\protect\citeauthoryear{{Pearson}, {Chamel}, {Goriely}  \&
						{Ducoin}}{{Pearson} et~al.}{2012}]{PCGD12}
					{Pearson} J.~M.,  {Chamel} N.,  {Goriely} S.,   {Ducoin} C.,  2012, \mn@doi
					[Phys. Rev. C] {10.1103/PhysRevC.85.065803}, \href
					{http://adsabs.harvard.edu/abs/2012PhRvC..85f5803P} {85, 065803}
					
					\bibitem[\protect\citeauthoryear{{Posselt}, {Pavlov}, {Suleimanov}  \&
						{Kargaltsev}}{{Posselt} et~al.}{2013}]{PPSK13}
					{Posselt} B.,  {Pavlov} G.~G.,  {Suleimanov} V.,   {Kargaltsev} O.,  2013,
					\mn@doi [ApJ] {10.1088/0004-637X/779/2/186}, \href
					{http://adsabs.harvard.edu/abs/2013ApJ...779..186P} {779, 186}
					
					\bibitem[\protect\citeauthoryear{{Potekhin}}{{Potekhin}}{2014}]{Potekhin_14}
					{Potekhin} A.~Y.,  2014, Phys.-Usp., \href
					{http://adsabs.harvard.edu/abs/2014arXiv1403.0074P} {57, 735}
					
					\bibitem[\protect\citeauthoryear{{Potekhin}, {Chabrier}  \&
						{Yakovlev}}{{Potekhin} et~al.}{1997}]{PCY97}
					{Potekhin} A.~Y.,  {Chabrier} G.,   {Yakovlev} D.~G.,  1997, A\&A, \href
					{http://adsabs.harvard.edu/abs/1997A%26A...323..415P} {323, 415}
						
						\bibitem[\protect\citeauthoryear{{Potekhin}, {Yakovlev}, {Chabrier}  \&
							{Gnedin}}{{Potekhin} et~al.}{2003}]{PYCG03}
						{Potekhin} A.~Y.,  {Yakovlev} D.~G.,  {Chabrier} G.,   {Gnedin} O.~Y.,  2003,
						\mn@doi [ApJ] {10.1086/376900}, \href
						{http://adsabs.harvard.edu/abs/2003ApJ...594..404P} {594, 404}
						
						\bibitem[\protect\citeauthoryear{{Potekhin}, {Fantina}, {Chamel}, {Pearson}  \&
							{Goriely}}{{Potekhin} et~al.}{2013}]{Potekhin_etal13}
						{Potekhin} A.~Y.,  {Fantina} A.~F.,  {Chamel} N.,  {Pearson} J.~M.,   {Goriely}
						S.,  2013, \mn@doi [A\&A] {10.1051/0004-6361/201321697}, \href
						{http://adsabs.harvard.edu/abs/2013A%26A...560A..48P} {560, A48}
							
							\bibitem[\protect\citeauthoryear{{Potekhin}, {Pons}  \& {Page}}{{Potekhin}
								et~al.}{2015}]{PPP15}
							{Potekhin} A.~Y.,  {Pons} J.~A.,   {Page} D.,  2015, \mn@doi [Space Sci. Rev.]
							{10.1007/s11214-015-0180-9}, \href
							{http://cdsads.u-strasbg.fr/abs/2015SSRv..191..239P} {191, 239}
							
							\bibitem[\protect\citeauthoryear{{Rosen}}{{Rosen}}{1968}]{Rosen68}
							{Rosen} L.~C.,  1968, \mn@doi [Ap\&SS] {10.1007/BF00656008}, \href
							{http://adsabs.harvard.edu/abs/1968Ap%26SS...1..372R} {1, 372}
								
								\bibitem[\protect\citeauthoryear{{Shternin} \& {Yakovlev}}{{Shternin} \&
									{Yakovlev}}{2015}]{SY15}
								{Shternin} P.~S.,  {Yakovlev} D.~G.,  2015, \mn@doi [MNRAS]
								{10.1093/mnras/stu2339}, \href
								{http://adsabs.harvard.edu/abs/2015MNRAS.446.3621S} {446, 3621}
								
								\bibitem[\protect\citeauthoryear{{Shternin}, {Yakovlev}, {Heinke}, {Ho}  \&
									{Patnaude}}{{Shternin} et~al.}{2011}]{Shternin_etal11}
								{Shternin} P.~S.,  {Yakovlev} D.~G.,  {Heinke} C.~O.,  {Ho} W.~C.~G.,
								{Patnaude} D.~J.,  2011, \mn@doi [MNRAS] {10.1111/j.1745-3933.2011.01015.x},
								\href {http://adsabs.harvard.edu/abs/2011MNRAS.412L.108S} {412, L108}
								
								\bibitem[\protect\citeauthoryear{{Weisskopf}, {Tennant}, {Yakovlev}, {Harding},
									{Zavlin}, {O'Dell}, {Elsner}  \& {Becker}}{{Weisskopf}
									et~al.}{2011}]{Weisskopf_etal11}
								{Weisskopf} M.~C.,  {Tennant} A.~F.,  {Yakovlev} D.~G.,  {Harding} A.,
								{Zavlin} V.~E.,  {O'Dell} S.~L.,  {Elsner} R.~F.,   {Becker} W.,  2011,
								\mn@doi [ApJ] {10.1088/0004-637X/743/2/139}, \href
								{http://adsabs.harvard.edu/abs/2011ApJ...743..139W} {743, 139}
								
								\bibitem[\protect\citeauthoryear{{Yakovlev} \& {Pethick}}{{Yakovlev} \&
									{Pethick}}{2004}]{YP04}
								{Yakovlev} D.~G.,  {Pethick} C.~J.,  2004, \mn@doi [ARA\&A]
								{10.1146/annurev.astro.42.053102.134013}, \href
								{http://adsabs.harvard.edu/abs/2004ARA%26A..42..169Y} {42, 169}
									
									\bibitem[\protect\citeauthoryear{{Yakovlev}, {Kaminker}, {Gnedin}  \&
										{Haensel}}{{Yakovlev} et~al.}{2001}]{YKGH01}
									{Yakovlev} D.~G.,  {Kaminker} A.~D.,  {Gnedin} O.~Y.,   {Haensel} P.,  2001,
									\mn@doi [Phys. Rep.] {10.1016/S0370-1573(00)00131-9}, \href
									{http://adsabs.harvard.edu/abs/2001PhR...354....1Y} {354, 1}
									
									\bibitem[\protect\citeauthoryear{{Yakovlev}, {Ho}, {Shternin}, {Heinke}  \&
										{Potekhin}}{{Yakovlev} et~al.}{2011}]{Yakovlev_etal11}
									{Yakovlev} D.~G.,  {Ho} W.~C.~G.,  {Shternin} P.~S.,  {Heinke} C.~O.,
									{Potekhin} A.~Y.,  2011, \mn@doi [MNRAS] {10.1111/j.1365-2966.2010.17827.x},
									\href {http://adsabs.harvard.edu/abs/2011MNRAS.411.1977Y} {411, 1977}
									
									\bibitem[\protect\citeauthoryear{{Zavlin}}{{Zavlin}}{2009}]{Zavlin09}
									{Zavlin} V.~E.,  2009, in {Becker} W.,  ed.,  Astrophysics and Space Science
									Library Vol. 357, Neutron Stars and Pulsars. p.~181
									
									\makeatother
								\end{thebibliography}
\end{document}